\documentclass{aa}  

\usepackage{graphicx}
\usepackage[varg]{txfonts} 
\usepackage{lipsum}
\usepackage{subcaption}         
\usepackage{lscape}             
\usepackage{placeins}           
\usepackage{dcolumn}
\usepackage[normalem]{ulem}

\newcolumntype{P}[1]{D{\pm}{\pm}{#1}}
\usepackage{color}

\usepackage[colorlinks=true, allcolors=blue]{hyperref}

\begin{document}
   \title{Ram-pressure-induced star formation in low-mass galaxies infalling on-to the Coma cluster: insights from DESI}
   \titlerunning{Ram-pressure-induced star formation in Coma in low-mass galaxies}

\author{Kirill A. Grishin\inst{1, 2}
        \and
        Igor V. Chilingarian\inst{3, 2}
        \and
        Gary A. Mamon\inst{4}
        \and
        Andrea Biviano\inst{5,6}
        \and
        Aleksandra Sharonova\inst{2,7}
        }

 \institute{Universite Paris Cite, CNRS, Astroparticule et Cosmologie, F-75013 Paris, France
            \and
            Sternberg Astronomical Institute, M.V. Lomonosov Moscow State University, 13 Universitetsky prospect, 119992 Moscow, Russia
            \and
            Center for Astrophysics | Harvard \& Smithsonian, 60 Garden St., Cambridge, MA 02138, USA
            \and
            Institut d'Astrophysique de Paris (UMR 7095: CNRS \& Sorbonne Université), 98 bis Bd Arago, 75014, Paris, France
            \and
            INAF Osservatorio Astronomico di Trieste, Via G.B. Tiepolo 11, 34143, Trieste, Italy
            \and
            IFPU: Institute for Fundamental Physics of the Universe, Via Beirut, 2, 34014, Italy
            \and
            Faculty of Space Research, M.V. Lomonosov Moscow State University, 1 Leninskiye Gory, 119991 Moscow, Russia
}
   \date{Received September 30, 20XX}
 
  \abstract{Ram-pressure stripping is a key driver of galaxy morphological transformation in clusters, contributing to the formation of quenched, especially dwarf, populations. Ram-pressure compression can also induce a starburst prior to quenching and build up significant stellar mass in an initially gas-rich galaxy. The detailed physics of these processes remains poorly understood, especially in the low-mass regime. Here we demonstrate that the key factor for a ram-pressure induced starburst in a low-mass galaxy is its angular momentum within a host cluster. 
  In this study, we select a sample of 41 post-starburst galaxies (PSGs) in the Coma cluster using the DESI EDR spectroscopic data, extending to low luminosities ($M_g < -14$). This sample is at least 90\% complete down to $M_g \approx -14.8$, which enabled us a systematic analysis of their properties. For each galaxy, we use projected cluster-centric distances and line-of-sight velocities to constrain the normalized orbital angular momentum and a 3D radial coordinate to the cluster center, assuming zero orbital energy. The resulting probability distributions show that while star-forming galaxies are split into two populations favoring intermediate and high angular momentum, almost all PSGs prefer high angular momentum. Our analysis statistically demonstrates that ram-pressure–induced starbursts are more efficient on tangential orbits, where gas stripping proceeds slowly enough to allow substantial star formation before gas removal.}


   \maketitle

\nolinenumbers

\section{Introduction}

Tracing galaxy formation and evolution is essential for understanding baryonic processes in the Universe \citep{2024MNRAS.532.3417W}. On galactic scales, observations remain one of the primary sources of information about structure growth in the strongly non-linear regime, particularly because simulation-based studies are often limited by resolution and rely on subgrid physics prescriptions that can introduce biases \citep{2015MNRAS.450.1937C, 2019MNRAS.490.3196P}.

Galaxy properties such as size, stellar and dynamical mass, color, and morphology encode the physical processes that shape their evolution. These processes can be broadly divided into two categories: (1) internal mechanisms, including star formation and active galactic nucleus (AGN) feedback \citep{2012ARA&A..50..455F, 2009ApJ...695..292C}, and (2) external processes, such as mergers \citep{1996ApJS..107....1A, 2012A&A...548A...7L}, gas accretion \citep{2002A&A...392...83B}, tidal stripping \citep{1983ApJ...264...24M}, galaxy ``harassment''~\citep{1996Natur.379..613M}, and ram-pressure stripping \citep{1972ApJ...176....1G}. Because environment-driven processes depend on local environmental properties, such as galaxy number density~\citep[e.g. ][]{1980ApJ...236..351D}, they produce correlations between galaxy populations and their environments. In particular, dense environments like massive galaxy clusters host a higher fraction of quiescent galaxies, indicating higher efficiency of the environmental mechanisms \citep{2006MNRAS.373..469B, 2007MNRAS.381....7H, 2026arXiv260206119M}, at the same time, in clusters, galaxies of different morphological types demonstrate different statistical properties of their orbits, such as orbital anisotropy~\citep{2019A&A...631A.131M}.

Galaxy clusters, as the largest gravitationally bound structures in the Universe, represent the most extreme environments for galaxy evolution and are highly efficient at quenching star formation and producing quiescent systems. In clusters one of the most effective transformation mechanisms is ram-pressure stripping \citep{2008MNRAS.383..593M}, in which the interaction between the hot intracluster medium (ICM) and the interstellar medium (ISM) of galaxies removes their gas reservoirs. The ram pressure experienced by a galaxy is given by $P_\mathrm{ ram} = \rho\, v^2$, where $\rho$ is the density of the hot gas and $v$ is the galaxy velocity within the cluster. As a result, the efficiency of ram-pressure stripping increases rapidly with cluster mass, since more galaxies in more massive clusters have higher velocities~\citep{2006ApJ...647..910H}.

Both observational and simulation-based studies identify ram-pressure stripping as the primary driver of the formation of quiescent galaxy populations in clusters \citep{2006PASP..118..517B, 2008ApJ...674..742B, 2008MNRAS.383..593M, 2022A&ARv..30....3B}. This process is particularly effective for dwarf galaxies, which dominate cluster populations numerically~\citep{2020MNRAS.494.1114S, 2021A&A...650A..99J}. Their shallow gravitational potential wells are less capable of retaining gas against ram pressure, as the restoring force, $F = 2 \pi\, G\, \Sigma_*\, \Sigma_\mathrm{ gas}$, is significantly smaller than in more massive galaxies. Despite its importance, the evolution of dwarf galaxies in clusters remains poorly captured in cosmological simulations due to limited resolution, which prevents reliable modeling of systems with stellar masses below $M_* \lesssim 10^8\,M_\odot$ \citep{2024MNRAS.532.1814G, 2024A&A...690A.286A}, and subgrid prescriptions do not fully allow to properly describe ISM–ICM interactions on sub-kiloparsec scales.

Ram pressure can also enhance star formation by compressing the interstellar medium~\citep{1985ApJ...294L..89G, 2009A&A...499...87K}. The efficiency of such ram-pressure–induced starbursts depends on multiple factors~\citep{2020MNRAS.494.1114S}, and in some cases they may account for up to 30\% of the total stellar mass formed in a galaxy \citep{2021NatAs...5.1308G}. Subsequent ram-pressure stripping can then remove the remaining gas, forcing the galaxy to evolve passively thereafter. The presence of a strong starburst immediately prior to quenching leads to systems that retain blue colors typical of star-forming galaxies but lack current star formation. These objects are classified as post-starburst galaxies (PSGs) or as ``k+a'' or ``E+A'' systems~\citep{1983ApJ...270....7D, 1987MNRAS.229..423C} given that spectra of such galaxies contain features typical for K-type stars and, hence, old populations similarly to elliptical galaxies, and features typical for A-type stars like deep Balmer absorption lines, indicators of a recent starburst. PSGs have been identified in significant numbers in nearby clusters such as Coma and Virgo, where they constitute a distinct subpopulation of cluster galaxies \citep{2004ApJ...601..197P}.

Despite their importance for galaxy evolution and stellar mass assembly, ram-pressure--induced starbursts (RP-induced SB) are poorly reproduced in cosmological simulations due to limited resolution, making it difficult to identify the galaxy properties that regulate their efficiency. Observational studies therefore provide critical constraints on the role of galaxy orientation and orbital parameters in triggering RP-induced SB and in shaping the PSG population in clusters.

Observations and simulations have shown that ram-pressure stripping drives the morphological transformation of low-mass, gas-rich galaxies into dwarf systems, such as dwarf ellipticals and ultra-diffuse galaxies (UDGs)~\citep{2022A&A...667A..76J}, often through a post-starburst galaxy (PSG) phase \citep{1999ApJ...518..576P, 2021NatAs...5.1308G}. Observations of PSGs in the Coma cluster further reveal populations of star clusters formed during this recent episode of star formation. At the same time, observations with space observatories have demonstrated that some UDGs host rich globular cluster systems~\citep{2025MNRAS.536.1217F} and massive nuclear star clusters. Detailed studies of PSGs, as the likely progenitors of different types of low-mass galaxies~\citep{2004ApJ...607..258Y, 2008ApJ...688..945Y}, including UDGs, can therefore constrain the physical conditions that regulate ram-pressure-induced star formation and potentially supply the building blocks for globular and nuclear star clusters in UDGs. Previous spectroscopic surveys, such as SDSS, were limited by small sample sizes, preventing robust statistical conclusions.

In this study, we use spectroscopic data from the DESI survey to identify PSGs in the Coma cluster, extending the sample to lower stellar masses and luminosities. The increased sample size together with a high completeness enables a statistical analysis of galaxy orbits and an assessment of how orbital parameters influence the efficiency of RP-induced SB.

Throughout this work, we adopt a distance to the Coma cluster of $101\, \mathrm{Mpc}$ \citep{2003MNRAS.343..401L}, corresponding to an angular scale of $0.49\, \mathrm{kpc\, arcsec^{-1}}$. We assume $H_0 = 70\, \mathrm{km\, s^{-1}\, Mpc^{-1}}$, an overdensity relative to the critical density of $\Delta_\mathrm{ c} = 102$, and take the cluster center to be in the middle between two massive cDs, NGC 4874 and NGC 4889 in projection, with a systemic velocity of $6900\, \mathrm{km\, s^{-1}}$ \citep{2003MNRAS.343..401L} and a distance modulus of $(m - M)=34.99$. All magnitudes are given in the AB system \citep{1983ApJ...266..713O}, and all quoted uncertainties correspond to the 1$\sigma$ confidence level unless stated otherwise.

\section{Data and data analysis}
\subsection{Hyper Suprime-Cam imaging data}
\label{sec:hsc_sepp}
\begin{figure*}
    \centering
    \includegraphics[width=\hsize]{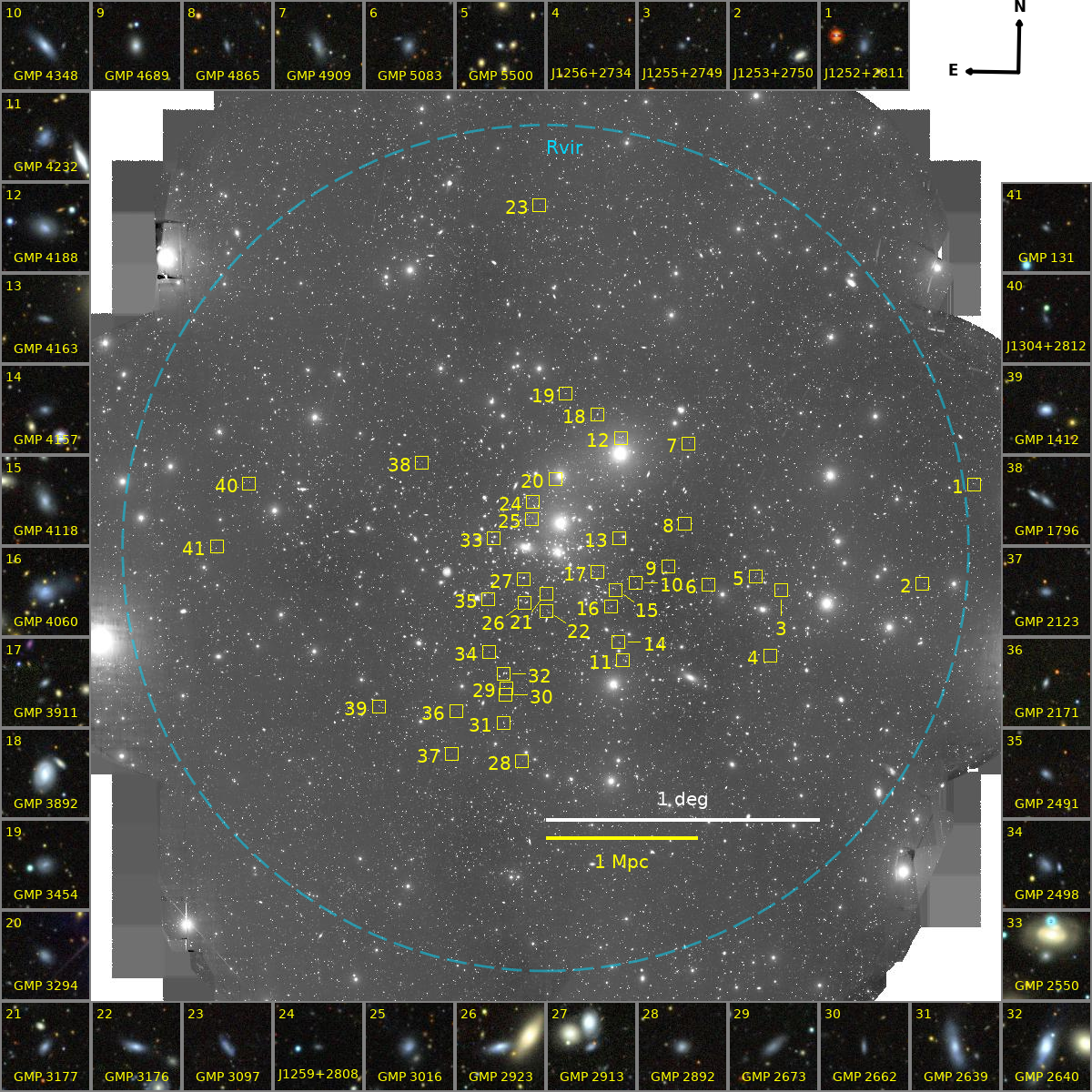}
    \caption{Position (yellow squares) of the post-starburst galaxies (PSGs) identified in Coma cluster with DESI EDR spectroscopic sample overlaid on a HSC $g$-band image and their color cutouts from Legacy Surveys with a name of a galaxy and a number in top left corner matching the yellow square on the map. Dashed cyan circle matches the virial radius of Coma ($R_\mathrm{vir}$) we adopted in this study.}\label{fig:coma_psg}
\end{figure*}

For galaxy shape measurements, we use deep imaging data of the Coma cluster obtained with the Hyper Suprime-Cam (HSC) instrument \citep{2018PASJ...70S...1M} in the $g$ band, primarily collected during 2016 and 2017 under seeing conditions ranging from 0.59 to 1.27 arcsec (Fig.~\ref{fig:coma_psg}). The dataset covers the cluster out to $1.23\,R_\mathrm{ vir}$, providing a unique opportunity to trace galaxy evolution in the cluster outskirts. Details of the data reduction and processing are presented in \citet{2025ApJ...993..229S}.
The depth of the HSC imaging, combined with accurate sky-background subtraction, enables robust measurements of galaxy structural parameters, including effective radii. In contrast, standard empirical background subtraction methods can oversubtract the low-surface-brightness outer regions of galaxies and lead to systematic underestimates of their sizes.

\subsection{DESI spectroscopic dataset}

The selection of PSGs in the Coma cluster must rely on spectroscopic redshift measurements, since galaxies of this type exhibit atypical colors. Due to the presence of a recent starburst, they appear significantly bluer than typical quenched systems. As a result, in the color--magnitude diagram~\citep{1958MeLuS.136....1H, 1961ApJS....5..233D} they lie below the red sequence and are  located in the ``green valley'', or even in regions typically populated by star-forming galaxies. Including a UV color (e.g. $\mathrm{NUV}-r$) does not fully resolve this issue. UV colors trace more recent star formation than optical colors and remain sensitive to quenching events that occurred $\sim$200--300 Myr ago~\citep{2012MNRAS.419.1727C}. Consequently, PSGs may still deviate from the locus of truly quiescent galaxies even in UV--optical color space. Spectroscopy-based selection is therefore essential. It allows the identification of fully quiescent systems with no residual star formation, which would otherwise produce weak $\rm{H\alpha}$ emission that cannot be reliably detected using broad-band photometric criteria alone. Even faint $\rm{H\alpha}$ emission is important, as its presence indicates that a galaxy is not fully quenched and may still experience future star formation.

In this study, we use spectroscopic data for the Coma cluster obtained by the Dark Energy Spectroscopic Instrument (DESI) survey \citep{2022AJ....164..207D} and released as part of the Early Data Release \citep[EDR;][]{2024AJ....168...58D}. A detailed description of the dataset and its composition can be found in \citet{2025ApJ...993..229S}. The sample includes spectra for 3292 galaxies in the Coma cluster and constitutes one of the most complete and homogeneous spectroscopic surveys of Coma cluster members to date.
Given the composition of a DESI spectroscopic sample, we expect a complete sample down to $M_r \approx -15.5$, since it is fully covered by the BGS Bright sample and a completeness at the level of at least $80-85\%$ for $M_r < -14.8$, as expected for the BGS Faint sample.

\subsubsection{DESI spectra modeling with Nbursts}

We model the DESI EDR spectra of Coma cluster members using the NBursts package for full spectral fitting \citep{2007MNRAS.376.1033C,2007IAUS..241..175C}, adopting X-shooter simple stellar population (SSP) templates \citep{2022A&A...661A..50V}. The fitting procedure and details of the modeling configuration are described in \citet{2025ApJ...993..229S}. As a result, for each galaxy we obtain estimates of stellar age and metallicity under the assumption of a single-burst stellar population model with fixed age and metallicity.

\section{Sample selection and properties}
\label{sec:sel_crit}

We select PSGs using criteria similar to those applied to the SDSS-based sample in \citet{2021NatAs...5.1308G}. However, objects observed with DESI are systematically fainter than in a SDSS sample, what results in lower signal-to-noise ratios -- and, consequently, less precise stellar population parameters -- we do not impose explicit constraints on stellar age or metallicity. The stellar age is instead indirectly constrained through the color-based selection.

We apply the following selection criteria to the stellar population parameters derived from DESI spectral modeling, complemented by photometric measurements from the Legacy Surveys \citep{2019AJ....157..168D}, which are on average 2.0--2.5 mag deeper than SDSS:

\begin{itemize}
\item $F_\mathrm{H\alpha} < 5\, \sigma(F_\mathrm{H\alpha})$, to select galaxies with no current star formation;
\item $g - r < 0.5$~mag, to identify objects with blue colors, indicating recent starburst;
\item $g < 21$~mag (integrated magnitude), to ensure sufficient continuum S/N.
\end{itemize}

The final criterion was introduced after visual inspection of the spectra selected by applying the first two conditions, as none of the objects with $m_g > 21$~mag exhibited a sufficient S/N ratio to reveal prominent spectral features such as absorption lines or the Balmer break and did not even allow us to conclude about the cluster membership.

Applying these criteria to the parent sample yields 46 candidate PSGs. Five objects were excluded after visual inspection, either because their spectra exhibit features typical of high-redshift quasars or because the S/N is too low to confirm the presence of any spectral features implied by the DESI pipeline models and conclude about their cluster membership. This selection results in a final sample of 41 galaxies. Of these, 8 overlap with the SDSS-based PSG sample of \citet{2021NatAs...5.1308G}; all objects previously identified with SDSS are recovered except GMP~3640, which was not observed by DESI. A typical value for $\sigma(F_\mathrm{H\alpha})\sim 0.5-1.5 \times 10^{-17} \mathrm{erg/s/cm^2}$ puts a lower limit for the star formation rate for these PSGs inside a DESI fiber of 1.5'' (diameter) or 0.75 kpc below $10^{-3} M_\mathrm{\odot}\, \mathrm{yr}^{-1}$~\citep{1998ARA&A..36..189K}.

\begin{figure}
    \centering
    \includegraphics[width=\hsize]{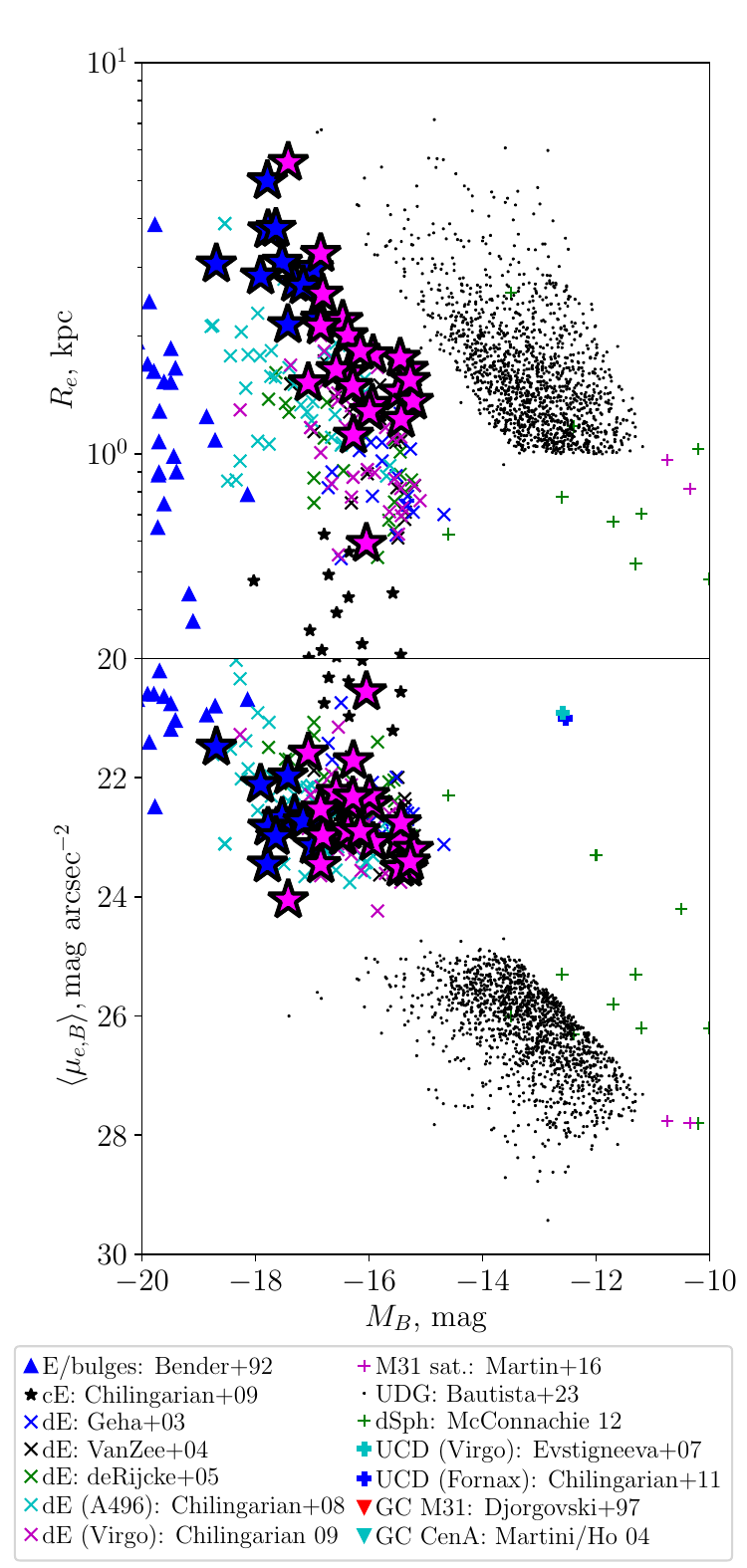}
    \caption{Size -- luminosity~\citep{1977ApJ...218..333K} and the surface brightness -- luminosity relations of early-type galaxies. With stars we show the position of the selected post-starburst galaxies, including those, previously identified with SDSS (\citet{2021NatAs...5.1308G}, blue stars) and new objects, identified with DESI (pink stars). For a reference here we show samples of objects from \citet{1992ApJ...399..462B, 2003AJ....126.1794G, 2005A&A...438..491D, Chilingarian+08, Chilingarian09, 1944ApJ...100..137B, yagi16, 2012AJ....144....4M, 2007AJ....133.1722E, Chilingarian+11, 1997ApJ...474L..19D,2004ApJ...610..233M, 2014MNRAS.443.1151N, 2008MNRAS.385L..83C}}
    \label{fig:muRe_Mv}
\end{figure}

The properties of the selected galaxies are summarized in Table~\ref{tab:gals}. Figure~\ref{fig:caustic_coma} and Figure~\ref{fig:muRe_Mv} show their locations on the projected phase-space diagram, size--luminosity and surface-brightness--luminosity relations, based on the results of the photometric decomposition (Sec.~\ref{sec:hsc_sepp}) respectively.
Compared to the SDSS-based PSG sample, the lower-luminosity PSGs identified with DESI are more compact and have effective radii that are on average a factor of 2--3 smaller. At the same time, their mean surface brightnesses are comparable to those of more luminous PSGs and dwarf ellipticals (dEs). In general, the newly identified PSGs span the parameter space between dEs, SDSS-selected PSGs, and UDGs, with effective radii $r_\mathrm{e} > 1$~kpc typical of UDGs and luminosities only 1--2~mag brighter. Taken together, these properties suggest that, under passive evolution similar to that of more luminous PSGs, these systems are likely to evolve into medium- and low-luminosity dEs/UDGs.

\subsection{Reference sample of low-mass star-forming galaxies}
\label{sec:sf_sel}

The moment when ram pressure induces a starburst and subsequently fully quenches a galaxy depend on several factors including the galaxy stellar mass, its orbit in a cluster and the properties of its ISM~\citep{2020MNRAS.494.1114S}. In low-mass dwarf galaxies, ram pressure is expected to be a particularly efficient mechanism driving the morphological transformation from actively star-forming systems to a quenched population. To trace this transformation, we selected a reference sample of actively star-forming galaxies (SFG) with $F_\mathrm{H\alpha} > 5\sigma_\mathrm{F_\mathrm{H\alpha}}$ which cover the same luminosity range as PSGs, $15.9 < m_\mathrm{r} < 21.0$~mag. On average, these galaxies reside at larger projected distances from the cluster center than PSGs, indicating that they have not yet reached the regions of the cluster where, due to higher $\rho(r)$ and $v$, quenching can proceed efficiently. Therefore, we can treat this sample as a probe of the initial state of the enviromentally-driven transformation of low-mass gas-rich galaxies in massive clusters. 

\begin{figure}
    \centering
    \includegraphics[width=\hsize]{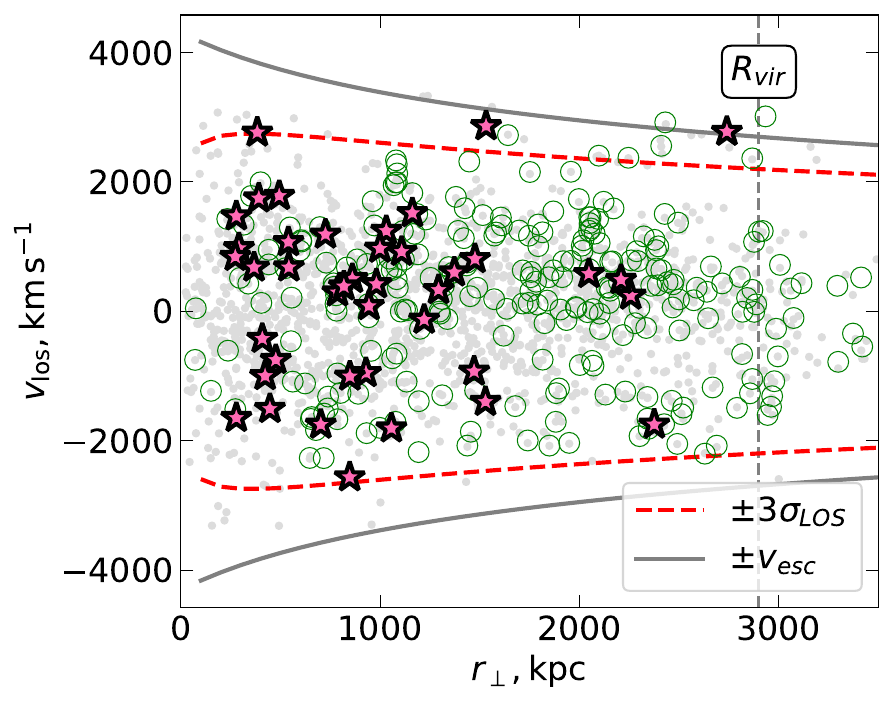}
    \caption{The projected phase space diagram of the Coma cluster members (grey points) based on DESI EDR redshift measurements with the position of the selected PSGs (magenta stars) and SFGs (open green circles). On this figure we also demonstrate a virial radius estimate, escape velocity, and line-of-sight velocity dispersion for the results of the modeling from \citet{2003MNRAS.343..401L} in a grey vertical dashed line, grey filled lines and red dashed lines respectively.}
    \label{fig:caustic_coma}
\end{figure}

In Fig.~\ref{fig:caustic_coma}, we also show the positions of the selected SFGs together with PSGs in the projected phase space. Both PSGs and SFGs demonstrate similar velocity dispersions and asymmetric velocity distribution with the respect to quiescent galaxies with higher probabilities for higher LOS velocities. This is consistent with the $v_\mathrm{los}$ asymmetry observed for blue galaxies in Coma~\citep{1996A&A...311...95B}, which can indicate the association of at least some infalling SFGs and PSGs with the NGC~4839 group, which hosts a population of galaxies with recent starbursts~\citep{1993AJ....106..473C}. However, a PSG closest to NGC~4839 is located 400~kpc (projected) away from it.

Given the low stellar masses of the identified SFGs, the most of these systems are likely to be stripped during their first orbit within the cluster and subsequently evolve either into PSGs for an extended period of time or, if the efficiency of the ram-pressure--induced star formation is low, almost directly into quiescent dwarf systems.

Therefore, the positions of PSGs and SFGs in the projected phase space allow us to trace ram-pressure--driven transformation processes in the low-mass regime, in which a galaxy with ongoing star formation evolves into a quiescent system either through a PSG phase lasting $\sim$1--2~Gyr or directly without a final episode of enhanced star formation.

\begin{table*}[]
    \centering
    \begin{tabular}{llcccclD{,}{\pm}{-1}D{,}{\pm}{-1}r}
    \hline
    \hline
    \# & Galaxy & RA & Dec. & $M_g$& $(g-r)$ &\multicolumn{1}{c}{$V$} & \multicolumn{1}{c}{{\rm Age}} & \multicolumn{1}{c}{\rm [Z/H]} & \multicolumn{1}{c}{$r_\mathrm{\perp}$}\\
    & & (deg) & (deg) & & & \multicolumn{1}{c}{($\rm{km\ s^{-1}}$)} & \multicolumn{1}{c}{($\rm{Myr}$)} & \multicolumn{1}{c}{(dex)} & (kpc)\\
    \hline

1 & J125239.5+281150 & 193.1646 & 28.1974 & $-$16.14 & 0.384\,$\pm$\,0.007 & 9733\,$\pm$\,155 & 349 , 74 & -0.64 , 0.59 & 2779\\
2 & J125332.6+275012 & 193.3860 & 27.8365 & $-$14.30 & 0.284\,$\pm$\,0.022 & 5105\,$\pm$\,42 & 2488 , 71 & -1.68 , 0.19 & 2422\\
3 & J125553.2+274903 & 193.9718 & 27.8174 & $-$14.61 & 0.488\,$\pm$\,0.023 & 5944\,$\pm$\,14 & 910 , 92 & 0.20 , 0.21 & 1532\\
4 & J125604.8+273444 & 194.0201 & 27.5790 & $-$14.23 & 0.343\,$\pm$\,0.021 & 9821\,$\pm$\,30 & 766 , 41 & -0.10\textsuperscript{*} & 1590\\
5 & GMP 5500 & 194.0785 & 27.8674 & $-$15.94 & 0.498\,$\pm$\,0.005 & 7232\,$\pm$\,5 & 1218 , 38 & -0.11 , 0.07 & 1356\\
6 & GMP 5083 & 194.2748 & 27.8392 & $-$16.45 & 0.389\,$\pm$\,0.007 & 7885\,$\pm$\,25 & 511 , 54 & -0.73 , 0.30 & 1067\\
7 & GMP 4909 & 194.3517 & 28.3581 & $-$16.91 & 0.458\,$\pm$\,0.005 & 7835\,$\pm$\,27 & 1276 , 160 & -0.51 , 0.35 & 1139\\
8 & GMP 4865 & 194.3695 & 28.0628 & $-$15.27 & 0.370\,$\pm$\,0.013 & 4272\,$\pm$\,24 & 814 , 70 & 0.20\textsuperscript{*} & 907\\
9 & GMP 4689 & 194.4388 & 27.9060 & $-$17.07 & 0.483\,$\pm$\,0.003 & 8109\,$\pm$\,7 & 1266 , 20 & -0.14 , 0.03 & 797\\
10 & GMP 4348 & 194.5759 & 27.8485 & $-$17.78 & 0.370\,$\pm$\,0.002 & 7588\,$\pm$\,4 & 750 , 25 & -0.75 , 0.06 & 617\\
11 & GMP 4232 & 194.6275 & 27.5643 & $-$16.93 & 0.337\,$\pm$\,0.005 & 7289\,$\pm$\,22 & 499 , 44 & -0.18 , 0.17 & 871\\
12 & GMP 4188 & 194.6347 & 28.3781 & $-$17.60 & 0.397\,$\pm$\,0.002 & 5864\,$\pm$\,13 & 462 , 20 & -0.66 , 0.15 & 855\\
13 & GMP 4163 & 194.6438 & 28.0098 & $-$15.45 & 0.459\,$\pm$\,0.008 & 6461\,$\pm$\,43 & 1363\textsuperscript{*} & -0.10 , 0.47 & 476\\
14 & GMP 4157 & 194.6464 & 27.6314 & $-$15.94 & 0.396\,$\pm$\,0.008 & 5104\,$\pm$\,87 & 616 , 112 & -0.94 , 0.46 & 759\\
15 & GMP 4118 & 194.6589 & 27.8225 & $-$17.42 & 0.459\,$\pm$\,0.003 & 5352\,$\pm$\,16 & 1456 , 82 & -0.35 , 0.05 & 521\\
16 & GMP 4060 & 194.6776 & 27.7605 & $-$18.00 & 0.301\,$\pm$\,0.002 & 8718\,$\pm$\,8 & 103 , 8 & -0.09 , 0.11 & 561\\
17 & GMP 3911 & 194.7312 & 27.8874 & $-$15.61 & 0.497\,$\pm$\,0.006 & 7893\,$\pm$\,11 & 2116 , 82 & -0.51 , 0.16 & 369\\
18 & GMP 3892 & 194.7338 & 28.4636 & $-$19.15 & 0.460\,$\pm$\,0.001 & 5919\,$\pm$\,2 & 1432 , 12 & -0.08 , 0.02 & 916\\
19 & GMP 3454 & 194.8641 & 28.5408 & $-$16.94 & 0.412\,$\pm$\,0.003 & 8165\,$\pm$\,15 & 639 , 89 & 0.06 , 0.23 & 997\\
20 & GMP 3294 & 194.9076 & 28.2281 & $-$16.28 & 0.420\,$\pm$\,0.004 & 6127\,$\pm$\,16 & 990 , 161 & -0.80 , 0.36 & 447\\
21 & GMP 3177 & 194.9425 & 27.8070 & $-$16.36 & 0.382\,$\pm$\,0.004 & 8397\,$\pm$\,13 & 1006 , 123 & -0.94 , 0.24 & 292\\
22 & GMP 3176 & 194.9429 & 27.7462 & $-$17.30 & 0.339\,$\pm$\,0.002 & 9721\,$\pm$\,8 & 550 , 14 & -0.79 , 0.08 & 398\\
23 & GMP 3097 & 194.9733 & 29.2332 & $-$16.47 & 0.336\,$\pm$\,0.005 & 7143\,$\pm$\,71 & 790 , 132 & 0.20\textsuperscript{*} & 2197\\
24 & J125959.8+280848 & 194.9993 & 28.1466 & $-$14.38 & 0.485\,$\pm$\,0.012 & 7587\,$\pm$\,13 & 1360 , 74 & -0.09\textsuperscript{*} & 310\\
25 & GMP 3016 & 195.0043 & 28.0821 & $-$16.89 & 0.349\,$\pm$\,0.002 & 7756\,$\pm$\,15 & 585 , 62 & 0.11 , 0.19 & 205\\
26 & GMP 2923 & 195.0335 & 27.7733 & $-$17.87 & 0.358\,$\pm$\,0.001 & 8679\,$\pm$\,4 & 244 , 9 & -0.57 , 0.10 & 373\\
27 & GMP 2913 & 195.0375 & 27.8610 & $-$16.04 & 0.395\,$\pm$\,0.004 & 5211\,$\pm$\,11 & 1490 , 58 & -0.32 , 0.20 & 239\\
28 & GMP 2892 & 195.0440 & 27.1944 & $-$16.58 & 0.496\,$\pm$\,0.005 & 7509\,$\pm$\,12 & 1202 , 95 & -0.09 , 0.14 & 1369\\
29 & GMP 2673 & 195.1094 & 27.4597 & $-$16.58 & 0.454\,$\pm$\,0.005 & 6977\,$\pm$\,21 & 1978 , 148 & -0.45 , 0.25 & 931\\
30 & GMP 2662 & 195.1138 & 27.4385 & $-$15.30 & 0.413\,$\pm$\,0.008 & 7323\,$\pm$\,27 & 979 , 111 & -0.23 , 0.12 & 968\\
31 & GMP 2639 & 195.1218 & 27.3332 & $-$18.38 & 0.357\,$\pm$\,0.002 & 8447\,$\pm$\,6 & 542 , 15 & -0.84 , 0.07 & 1149\\
32 & GMP 2640 & 195.1218 & 27.5148 & $-$19.14 & 0.340\,$\pm$\,0.001 & 7407\,$\pm$\,2 & 479 , 3 & -0.69 , 0.02 & 844\\
33 & GMP 2550 & 195.1630 & 28.0099 & $-$16.01 & 0.466\,$\pm$\,0.005 & 5872\,$\pm$\,8 & 1304 , 44 & -0.15 , 0.04 & 334\\
34 & GMP 2498 & 195.1833 & 27.5937 & $-$16.56 & 0.403\,$\pm$\,0.004 & 7195\,$\pm$\,28 & 394 , 64 & -0.11 , 0.43 & 755\\
35 & GMP 2491 & 195.1846 & 27.7887 & $-$15.94 & 0.380\,$\pm$\,0.006 & 7989\,$\pm$\,62 & 1051 , 134 & -0.59 , 0.43 & 485\\
36 & GMP 2171 & 195.3171 & 27.3774 & $-$15.64 & 0.499\,$\pm$\,0.005 & 6762\,$\pm$\,9 & 4568 , 693 & -1.06 , 0.18 & 1185\\
37 & GMP 2123 & 195.3359 & 27.2207 & $-$15.78 & 0.394\,$\pm$\,0.008 & 7721\,$\pm$\,31 & 585 , 88 & -0.88 , 0.34 & 1443\\
38 & GMP 1796 & 195.4630 & 28.2889 & $-$16.08 & 0.416\,$\pm$\,0.004 & 5045\,$\pm$\,14 & 590 , 37 & 0.14 , 0.14 & 962\\
39 & GMP 1412 & 195.6357 & 27.3934 & $-$17.51 & 0.353\,$\pm$\,0.002 & 5464\,$\pm$\,6 & 187 , 24 & -0.41 , 0.14 & 1464\\
40 & J130443.1+281228 & 196.1797 & 28.2078 & $-$14.77 & 0.327\,$\pm$\,0.013 & 7485\,$\pm$\,21 & 674 , 49 & 0.14 , 0.19 & 1938\\
41 & GMP 131 & 196.3120 & 27.9744 & $-$14.72 & 0.495\,$\pm$\,0.009 & 7392\,$\pm$\,8 & 1182 , 65 & -0.10 , 0.11 & 2099\\

\hline
\hline

        \end{tabular}
    \caption{Properties of post-starburst galaxies identified in Coma cluster with DESI EDR dataset.}
    \label{tab:gals}
    \tablefoot{Columns are: 1. galaxy number in a current table and in Fig.~\ref{fig:coma_psg}; 2. galaxy name from~\citet{1983MNRAS.202..113G} or, if absent, IAU name; 3. right ascension; 4. declination; 5. absolute magnitude in $g$-band; 6. integrated color $g-r$; 7. line-of-sight velocity $V$; 8. age of SSP model; 9. Metallicity $[Z/H]$ of the SSP model from the DESI spectrum; 10. projected distance to the cluster center. With \textsuperscript{*} we mark measurements that we consider unreliable due to insufficient S/N or limitations of the stellar population models.
    }

\end{table*}

\section{Constraints on the galaxy angular momentum}
\label{sec:Lminmax}

Statistical studies \citep{2004A&A...424..779B, 2009A&A...501..419B, 2014A&A...566A..68M, 2019A&A...631A.131M} already showed that different populations of galaxies demonstrate different properties of orbits within host clusters, first of all, velocity anisotropy $\beta$, which indicates how eccentric orbits are at the statistical level~\citep{1987ApJ...313..121M}, i.e. $\beta$ serves as a proxy for angular momentum, but it does not allow us to draw any conclusions for individual cases.
In general, the full six-dimensional phase-space position of an individual cluster member cannot be constrained observationally, because observations usually provide direct access to only three degrees of freedom: two spatial coordinates on the sky and a single component of the velocity vector along the line of sight. Additional constraints are therefore required to place even joint limits on the remaining parameters even for spherical systems. 

In a stationary, centrally symmetric slowly evolving gravitational potential $\Phi(r)$, the motion of a galaxy can be parametrized by its total orbital energy ($E$) and angular momentum vector ($\vec{L}$), both of which are conserved quantities in such systems.

For galaxies which infall into a massive cluster such as Coma, an assumption about the zero total orbital energy is a reasonable approximation, because any initial velocities are negligible compared to the cluster free-fall velocity ($v_\mathrm{\mathrm{ff}}$). This conclusion relies on the assumption that galaxies do not undergo strong interactions with neighbors, which is justified by the low galaxy density in the outer regions of the cluster, where most PSGs and SFGs are observed, and by the high velocity dispersion in clusters, which limits the energy exchange due to the short timescales of interactions. This constraint on $E$ allows us to explore the allowed range of the remaining integral of motion, $\vec{L}$, as a function of the only unconstrained spatial coordinate, line-of-sight depth, $z$. However, in a centrally symmetric potential, a constraint on $z$ alone does not contain additional physical information, since it is not invariant under geometric transformations such as rotation. In such systems, the physically relevant quantity describing a galaxy location within the cluster is a 3D distance from the cluster center, $R$, rather than an individual Cartesian coordinate.

\subsection{Estimation of normalized dimensionless angular momentum}

In the case of a centrally symmetric potential, one can apply a coordinate transformation only involving a rotation of the axes such that only a single component of $\vec{L}$ is non-zero. This allows the angular momentum to be described by a single scalar quantity, $L$, rather than by three vector components. Statistically, in a symmetric galaxy cluster, the orientation of the orbital plane has no effect on galaxy evolution for fixed values of $E$ and $L$, given the central symmetry of the system. The orbital properties such as the pericentric and apocentric distances and the orbital period, are therefore fully determined by the scalar pair $(E, L)$.

While the value of $L$ quantifies the degree of tangential motion, direct comparisons between individual objects and different galaxy populations can be significantly simplified by introducing a dimensionless analogue of $L$. This quantity can be defined as the absolute value of $\vec{L}$ normalized by the maximum possible angular momentum for a given $E$ and $r$ and can be written in two following forms:

\begin{equation}
\hat{L}(\vec{r},\vec{v}) = \frac{|\vec{L}|}{L_\mathrm{\max}(E,r)}
   = \frac{| \vec{r} \times \vec{v}   |}{r\,v}
   = |\sin\psi| 
= \sqrt{1 - \left(\frac{\vec{r}\cdot\vec{v}}{r\,v}\right)^2} \ ,
\label{eq:Ldef}
\end{equation}
where $L_\mathrm{\max}(E, r) = r\,v$ denotes the scalar (dot) product. This quantity directly corresponds to the absolute value of the sine of the angle between the radius and velocity vectors ($\psi$). 
Finally, given our observables $x$, $y$, and $v_z$, together with the assumption of $E = 0$, we can place constraints on the normalized angular momentum $\hat{L}$ of an individual cluster galaxy at a given 3D distance from the cluster center, $r$. 

The second form of expression for $\hat{L}(\vec{r},\vec{v})$ in Eq.~\ref{eq:Ldef} is the most convenient to analyze its minimum and maximum possible values.
For a given choice of total orbital energy $E$ and gravitational potential $\Phi(r)$, and treating $r$ as a free parameter, the extrema of $\hat{L}(\vec{r}, \vec{v})$ can be obtained by analyzing the extrema of the scalar product $(\vec{r}, \vec{v})$, which can be evaluated explicitly by expressing it through the components of $\vec{r}$ and $\vec{v}$ in a chosen coordinate system. This analysis yields the following expressions for the minimum and maximum values of $\hat{L}$ at a given $r$:

\begin{equation}
    \hat{L}_\mathrm{\substack{\min \\ \max^*}}(r) = \sqrt{1 - \frac{\left(v_\mathrm{pos}(r, v_z)\, r_\mathrm{\perp} \pm |v_\mathrm{los}|\sqrt{r^2 - r_\mathrm{\perp}^2}\right)^2}{r^2 \,   (v_\mathrm{pos}^2(r, v_z) + v_\mathrm{los}^2)}}
    \label{eq:Lminmax}
\end{equation}
Where 
\begin{equation}
    v_\mathrm{pos}^2(r, v_z) = v_x^2 + v_y^2 = 2[E - \Phi(r)] - v_\mathrm{los}^2 
\end{equation}

\noindent
is a plane-of-the-sky velocity and $r_\mathrm{\perp} = \sqrt{x^2 + y^2}$ is the projected distance to the cluster center. Here, $\hat{L}_\mathrm{\max}$ is conditional: the expression is valid only when $v_\mathrm{pos}(r, v_z) \cdot r_\mathrm{\perp} < |v_z| \sqrt{r^2 - r_\mathrm{\perp}^2}$; otherwise, $\hat{L}_\mathrm{\max} = 1$.

In Appendix~\ref{sec:Lstab} we discuss the robustness of this methodology against uncertainties and variations of the model parameters.

\subsection{Angular momentum probability distribution}
\label{sec:L_prob}

To derive the distributions of the normalized angular momentum for infalling galaxies, one must consider not only the allowed ranges of $\hat{L}$ but also the probability distribution of the 3D distance from the cluster center, $r$, given the observed projected distance $r_\mathrm{\perp}$. This distribution is described by the Abel transform of the three-dimensional galaxy number density profile, $\rho(r)$:

\begin{equation}
    f(r \mid r_\mathrm{\perp}) = 2\,\frac{\rho(r)}{\Sigma(r)}\,\frac{r}{\sqrt{r^2 - r^2_\mathrm{\perp}}}
\end{equation}
where $\Sigma(r)$ is a surface density profile.
If the distribution of galaxies traces the dark matter halo of a cluster, described by the NFW profile~\citep{2004ApJ...617..879L, 2005ApJ...633..122H, 2007A&A...461..411P}, the expression for $f(r \mid r_\mathrm{\perp})$ would be:
\begin{equation}
    f(r \mid r_\mathrm{\perp}) \propto \frac{1}{ (r + r_s)^2\sqrt{r^2 - r_\mathrm{\perp}^2}}
\end{equation}
where $r_s$ is the scale radius of the NFW profile. For an NFW profile, $f(r \mid r_\mathrm{\perp})$ cannot be normalized over the entire infinite range, so it is necessary to choose a maximum distance up to which the system is considered, which typically set to $R_\mathrm{200}$~\citep{2014MNRAS.441.1513O}.

\begin{figure}
    \centering
    \includegraphics[width=\hsize]{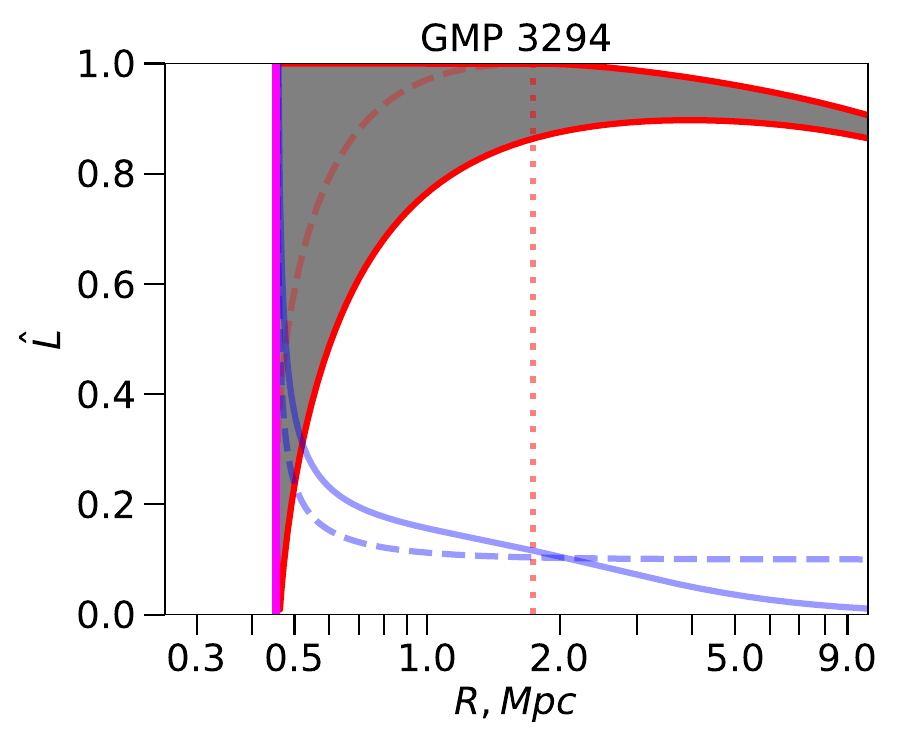}
    \caption{Assessment of the normalized total angular momentum ($\hat{L}$) and total cluster-centric distance ($r$) for the PSG GMP~3294. The grey-shaded region indicates the allowed values of $\hat{L}$ and $r$ based on the observed line-of-sight velocity ($v_z$) and projected distance to the cluster center ($r_\mathrm{\perp}$ = 0.447 Mpc), with the latter shown as a magenta line representing the minimal possible distance to the cluster center. The red dotted line marks the maximum 3D radius ($r$) for which $\hat{L}_\mathrm{max}$ = 1. The dashed and solid blue lines correspond to the Abel transform kernel multiplied by 0.1 and $f(r \mid d_\mathrm{\rm proj})$ for an NFW profile, respectively.\label{fig:Lr_exmpl}}
\end{figure}

The resulting probability distribution for $\hat{L}$ can be obtained by integrating the product of the two probability distributions, $f(r \mid r_\mathrm{\perp})$ and $f(L \mid r)$:
\begin{equation}
    f(\hat{L}) = \int{f(\hat{L} \mid r) \,f(r \mid r_\mathrm{\perp})\, \mathrm{d}r}
\end{equation}
where, under assumption of a flat prior on $\hat{L}$, $f(\hat{L} \mid r)$ can be written as follows:
\[
f(\hat{L} \mid r) = 
\begin{cases} 
\dfrac{1}{\hat{L}_\mathrm{\max}(r) - \hat{L}_\mathrm{\min}(r)}, & \hat{L}_\mathrm{\min}(r) < \hat{L} < \hat{L}_\mathrm{\max}(r) \\[2mm]
0, & \text{otherwise}
\end{cases}
\]
The integration of these expressions presents numerical challenges due to the singularity at \(r = r_\mathrm{\perp}\), which can introduce unexpected numerical effects. To address this, we first sample \(f(r \mid r_\mathrm{\perp})\) using a finite number of bins based on the integrated probability within each bin. For each sampled \(r\) value, we then sample the distribution of \(\hat{L}\). Finally, we combine all sampled distributions of \(\hat{L}\) and bin them to obtain the overall distribution. On Fig.~\ref{fig:Lr_exmpl} we show an example of constraints on \(\hat{L}\) and \(r\) obtained for an individual galaxy using Eq.~\ref{eq:Lminmax} and $f(r \mid r_\mathrm{\perp})$ of NFW profile for the Coma cluster.

\begin{figure*}
    \centering
    \includegraphics[width=0.49\hsize]{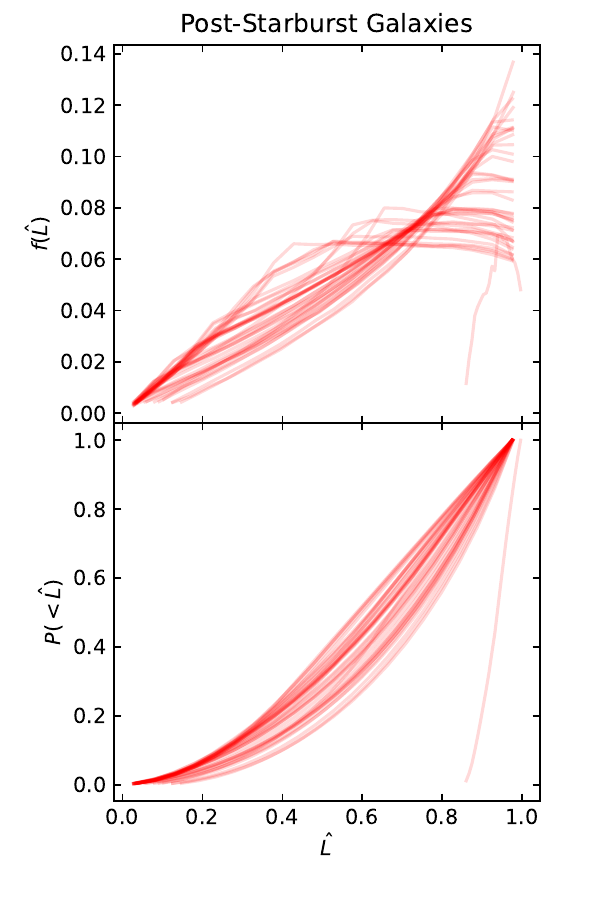}
    \includegraphics[width=0.49\hsize]{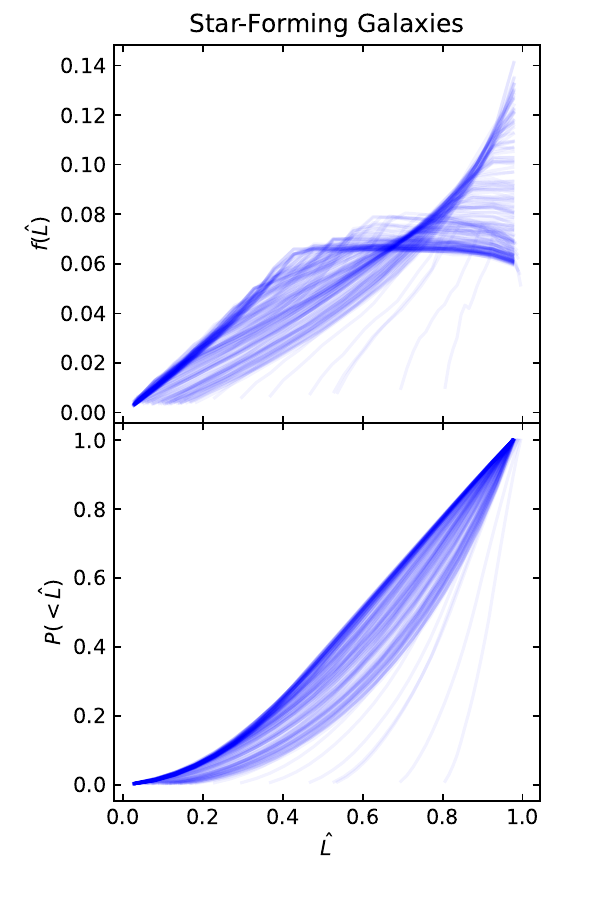}

    \caption{Individual probability distribution functions (PDF, upper panels) and cumulative distribution functions (CDF, lower panels) for $\hat{L}$ values from 0 to 1 for the sample of PSGs (left) and SFGs (right), each line is a different galaxy.}\label{fig:pL_psg}
\end{figure*}

\section{Discussion}

\subsection{Ram-pressure-induced starburst}

A starburst triggered by the ram-pressure compression can significantly affect a galaxy color, making it bluer due to the increased fraction of young stars, given that the most massive (and therefore bluest) stars are still on the main sequence. The fraction of stars formed during such an episode depends on several factors, primarily on the rate of ram-pressure growth during infall. If the ram pressure increases too rapidly, the gas is removed before it can be efficiently converted into stars. The full derivative of the ram pressure, \(P_\mathrm{\rm ram} = \rho(r)\, v^2(r)\), over the time is given by:

\begin{equation}\label{eq:dPdt}
    \begin{split}
    \frac{\mathrm{d}P_\mathrm{ram}}{\mathrm{d}t} & = v_r\left(v^2\frac{\mathrm{d}\rho(r)}{\mathrm{d}r} - 2\rho(r)\frac{\mathrm{d}\Phi(r)}{\mathrm{d}r}\right) \\ & = \rho v_r\frac{v^2}{r} \left[\frac{\mathrm{d}\ln \rho}{\mathrm{d}\ln r}+\frac{\mathrm{d}\ln(-\Phi)}{\mathrm{d}\ln r} \right].
    \end{split} 
\end{equation}

\noindent
Here, $\rho(r)$ is the density of the intracluster medium, \(\Phi(r)\) is the gravitational potential of the cluster, and \(v\) and \(v_r\) are the total velocity and its radial component, respectively. In this expression, \(v_r\) acts as a scale parameter for the rate of ram-pressure growth, indicating how rapidly and efficiently a galaxy can be stripped at a given location and along its velocity vector.  

Since the total squared velocity of a galaxy is composed of radial and tangential components, ($v^2 = v_r^2 + v_\mathrm{\perp}^2$) the tangential component ($v_\mathrm{\perp}$) also reflects the stripping rate. Unlike $v_r$, however, $v_\mathrm{\perp}$ is directly related to a conserved quantity that controls the galaxy motion in the cluster, namely its total angular momentum, ($L = r\,v_\mathrm{\perp}$), making it a more suitable parameter for characterizing the rate of the ram-pressure growth.

The very presence of a recent starburst in a galaxy star formation history can play a crucial role in its morphological transformation, affecting its total luminosity, color, stellar mass, and other properties. Because our selection criteria (Sec.~\ref{sec:sel_crit}) are based on color, we investigated how both the presence of a recent starburst and the fraction of stars formed in it influence a galaxy classification as a PSG. 

To this end, we studied the color evolution for a constant star formation history (SFH) followed by a starburst and subsequent quenching~\citep{2019arXiv190913460G}, generated using {\sc PEGASE} models~\citep{2004A&A...425..881L}. These templates have been shown to accurately reproduce both the spectra and spectral energy distributions of recently quenched PSGs~\citep{2021NatAs...5.1308G}.
Analysis of the PSG spectra and SED also showed that they are stable against variations of some parameters; e.g., deviations from a constant star-formation rate (SFR) before the final starburst do not affect the modeling results significantly thanks to the presence of the intense starburst in a galaxy SFH. This kind of averaging, when we assume a constant SFR as presented in our approach, does not have a dominating effect on the estimates of SFH parameters.

\begin{figure}
    \centering
    \includegraphics[width=\hsize]{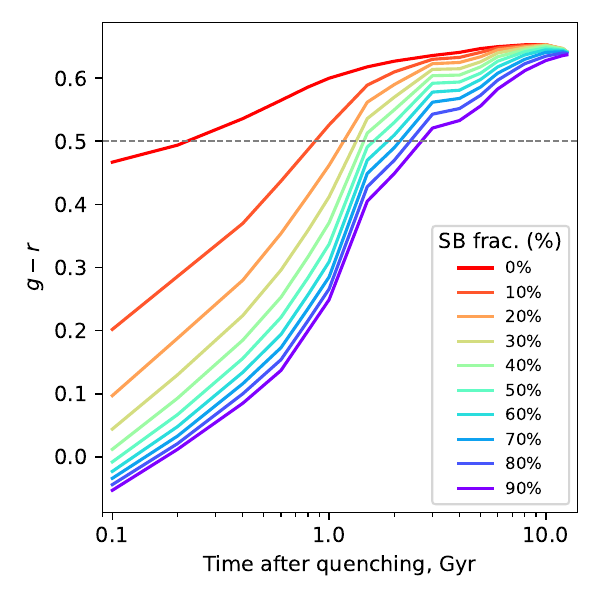}
    \caption{Evolution of galaxy $g - r$ color for different star formation histories (continuous plus final starburst (SB)). The tracks show how the color changes depending on the fraction of the total stellar mass that is formed during the SB relative to the entire SFH. The curves are calculated from a grid of the template spectral energy distributions from~\citet{2019arXiv190913460G}. A dashed gray line shows the color criterion used for the post-starburst selection.}\label{fig:col_rpssb}
\end{figure}

Figure~\ref{fig:col_rpssb} shows the evolution of the \(g-r\) color following a final quasi-instantaneous starburst, immediately followed by full  quenching, for a case where 99\% of the gas is converted into stars over the entire evolution of galaxies, a typical value for PSGs of lower mass from~\citet{2021NatAs...5.1308G}.
This figure illustrates that the presence of a recent starburst significantly impacts a galaxy color. For a truncated SFH (0\% starburst), the galaxy meets the PSG criteria for only the first 200 Myr. Since the color evolves very slowly, e.g., \(g-r = 0.46\,\mathrm{mag}\) after 100~Myr -- small variations in intrinsic properties, such as metallicity, can make the galaxy redder and prevent it from being classified as a PSG according to our criteria at any moment after quenching. In contrast, for a starburst contributing just 10\% of the total stellar mass, the galaxy maintains \(g-r < 0.5\,\mathrm{mag}\) for up-to 700~Myr after quenching, ensuring its classification as a PSG. This aligns well with the stellar population measurements for PSGs: in the main sample of 11 massive Coma PSGs from~\citet{2021NatAs...5.1308G}, 8 have the measured times after truncation (truncation ages) below 700~Myr.

\subsection{Post-starburst galaxies}
\label{sec:psg_l}
Using the methodology outlined in Sec.~\ref{sec:Lminmax}--\ref{sec:L_prob}, we derived the probability distributions (PDF) and cumulative distribution functions (CDF) for the selected PSGs. In our calculations, we adopted an NFW profile for the mass distribution of the Coma cluster, using the parameters from~\citet{2003MNRAS.343..401L}. The resulting distributions are shown in the left panels of Fig.~\ref{fig:pL_psg}. 

On average, PSGs tend to have high angular momentum resulting into the highest value of probability density in the range \(0.8 < \hat{L} < 1.0\), while the mean probability of \(\hat{L} < 0.5\) is 25--30\% (Fig.~\ref{fig:pL_psg}). This corresponds to large angles between the radius and velocity vectors,  \(50^\circ\)–\(130^\circ\), indicating that these galaxies move on highly tangential orbits. This behavior is consistent with the orientation of the tails of stripped material observed in more massive Coma PSGs~\citep{2021NatAs...5.1308G}, none of which point away from the cluster center, as would be expected for radial orbits.

\subsection{Star-forming galaxies}

For the subpopulation of Coma star-forming galaxies (SFGs; Sec.~\ref{sec:sf_sel}), we performed the same analysis as for PSGs described in Sec.~\ref{sec:psg_l}, with results shown in the right panels of Fig.~\ref{fig:pL_psg}. SFGs exhibit a clearer bimodality in their probability distributions for \(\hat{L}\), with a higher fraction of objects at lower \(\hat{L}\), where the probability distribution for \(\hat{L}\) reaches a plateau around \(\hat{L} \approx 0.4\), after which it remains roughly constant or even slightly decreases.

The cumulative distribution function (CDF), \(P(< \hat{L})\), which gives the probability of \(\hat{L}\) being smaller than a given value, highlights the difference between SFGs and PSGs more clearly. Around \(\hat{L} \approx 0.5\text{--}0.6\), the number of SFGs with high probability increases sharply, whereas for PSGs the CDF remains nearly constant.

While some SFGs exhibit a distribution function similar to that of PSGs, the higher fraction of low-\(\hat{L}\) cases results in an overall distribution that differs from that of PSGs.

\subsection{Statistical analysis of the PSGs and SFGs sample composition}

\begin{figure}
    \centering
    \includegraphics[width=\hsize]{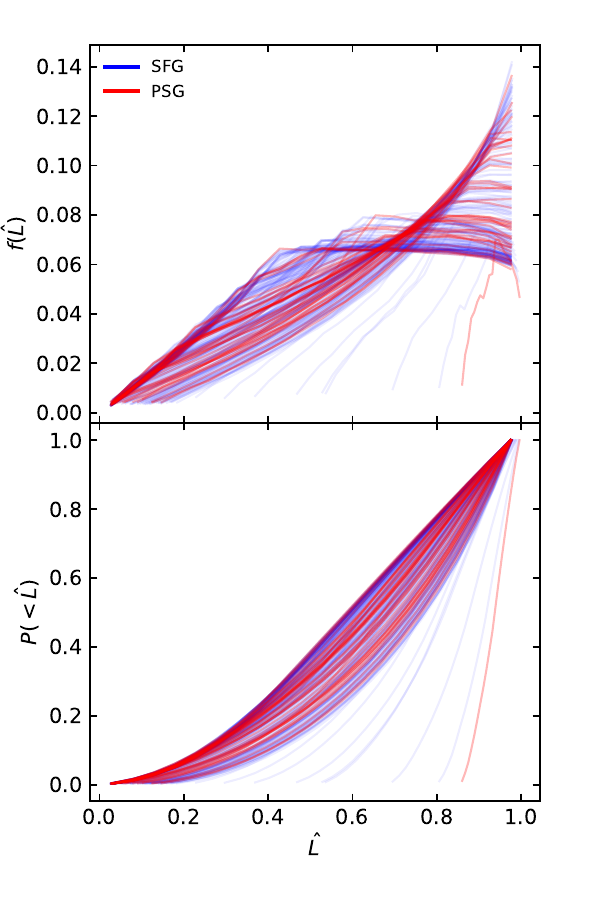}
    \caption{The PDF (upper panel) and CDF (lower panel) of PSGs (semi-transparent red) are overplotted on those of SFGs (semi-transparent blue). The dominance of the blue curves at high $p(\hat{L})$ values in the range $0.5 < \hat{L} < 0.6$ and at low $p(\hat{L})$ values in the range $0.9 < \hat{L} < 1.0$ indicates that SFGs contain a higher fraction of lower-$\hat{L}$ galaxies compared to PSGs.}\label{fig:L_comp}
\end{figure}

Both the PDF and CDF for SFGs and PSGs suggest that the two samples have different compositions. The SFG sample exhibits a larger fraction of systems with lower mathematical expectation of $\hat{L}$, in addition to a high-$\hat{L}$ subsample, which is more similar to PSGs (Fig.~\ref{fig:L_comp}).

Direct statistical tests comparing the probability distributions of SFGs and PSGs may not be sufficiently sensitive to differences of this kind. In particular, in a case of a comparison with a sample in which 65\% of the objects share the same distribution, while for the remaining 35\% the distributions differ only at the level of the third moment, since the mathematical expectations and standard deviations of both the low- and high-$\hat{L}$ subsamples are similar, the distinction between these object types appears primarily in the skewness, making it difficult for standard tests to detect such differences.

To properly account for the complexities discussed above, we performed two separate statistical comparisons: one between PSGs and high-$\hat{L}$ SFGs, and another one between PSGs and low-$\hat{L}$ SFGs. For this purpose, we divided the SFG sample into two subsamples according to the mean value of the PDF in the range $0.50 < \hat{L} < 0.55$. If this mean value exceeds 0.06, the galaxy is classified as low-$\hat{L}$; otherwise, it is classified as high-$\hat{L}$. Using this criterion, we obtained a high-$\hat{L}$ subsample that contains 116 galaxies and a low-$\hat{L}$ subsample that includes 65 galaxies. If applied to the PSG sample, this criterion classifies 90\% of them as high-$\hat{L}$ galaxies.

To quantify the differences between the two SFG subsamples and the PSGs, we generated 1000 realizations of $\hat{L}$ for each galaxy (for both SFG subsamples and PSGs) by randomly sampling from their respective PDFs. Therefore in each realization, the sample size was equal to the number of galaxies in the corresponding subsample. We then applied the Mann--Whitney $U$ rank test~\citep{MannWhitney47} to assess whether galaxies in each SFG subsample systematically exhibit lower or higher $\hat{L}$ values compared to PSGs, or whether the samples are statistically similar. This test provides the significance level, $p_\mathrm{sup}$, that a sample $A$ has systematically larger values than a sample $B$ based on the $U$ metrics, normalized by the product of the two sample sizes. If $p_\mathrm{sup} > 0.5$, we interpret that particular realization of $\hat{L}$ as an indication of systematically higher values in the sample $A$ relative to the sample $B$.

The statistical comparison between PSGs and the low-$\hat{L}$ SFG subsample yields $p_\mathrm{sup} > 0.5$ in 89\%\ of the realizations. This fraction can be interpreted as the significance level supporting the hypothesis that the selected low-$\hat{L}$ SFGs indeed have systematically lower $\hat{L}$ values than PSGs. In contrast, applying the same test to the high-$\hat{L}$ SFGs and PSGs results in $p_\mathrm{sup} > 0.5$ for only 38\%\ of the realizations. Since this value is much closer to 50\%, it indicates that the high-$\hat{L}$ SFG and PSG samples are statistically more similar in terms of their $\hat{L}$ distributions.

These results suggest that the SFG sample is composed of two statistically distinct components: a low-$\hat{L}$ subset, comprising 35\%\ of all SFGs, which differs significantly from the PSG population, and a high-$\hat{L}$ subset, which is statistically very similar to PSGs.

\section{Implications for a galaxy morphological transformation driven by ram pressure}
Both observations~\citep{2021PASP..133g2001F, 2022ApJ...930...43W} and simulations~\citep{2021MNRAS.506.4516L} indicate that ram-pressure stripping is the dominant mechanism responsible for the formation of PSGs in clusters. Ram pressure can quench a galaxy on relatively short timescales, typically less than half a billion years~\citep{2022ApJ...930...43W}, and even faster for low-mass systems. During this process, the stripped gas forms a tail behind the galaxy, producing the phenomenon commonly referred to as a ``jellyfish galaxy''~\citep{2010MNRAS.408.1417S}.  

Given the strong evolutionary connection between PSGs and jellyfish galaxies, the orbital properties of the latter provide valuable insights into the transformation pathways of galaxies into PSGs. Observational statistical studies of jellyfish galaxies based on their kinematics in their host cluster indicate that they predominantly move on radial orbits, exhibiting higher orbital anisotropy \(\beta\), which increases with distance from the cluster center~\citep{2018MNRAS.476.4753J, 2024ApJ...965..117B}.  

The orientation of the tail of stripped material relative to the cluster center also encodes the information about the orbit, under the assumption that systematic motions of the ICM are negligible. Systematic studies of tail directions reveal a wide range of orientations, with the most probable direction pointing away from the cluster center, as expected for radial orbits~\citep{2024MNRAS.533..341S}. However, observations do not completely rule out tails pointing perpendicular to, or even toward, the cluster center. Some studies suggest nearly equal probabilities for ``away'' and ``toward'' orientations~\citep{2022ApJ...934...86S}.  

The position along a galaxy orbit where stripping occurs also depends on its stellar mass. More massive galaxies can be stripped only in the central regions of the cluster, where the ICM density is high. For such galaxies, even in clusters as large as Coma, ram pressure may not fully strip the gas content during the first passage; such galaxies may experience the ``jellyfish'' phase over multiple passages.

The presence of gas tails not pointing away from the cluster center in massive galaxies can be explained if ram-pressure stripping occurs near the pericenter of the orbit. However, a similar situation is observed in lower-mass galaxies as well. For example, in the Coma cluster, GMP~4060 and GMP~2923~\citep[$M_* < 10^9 \,M_\mathrm{\odot}$, ][]{2021NatAs...5.1308G}, which are also included in our sample, exhibit tails perpendicular to the direction of the cluster center~\citep{2008ApJ...688..918Y, 2010MNRAS.408.1417S}.  If these two low-mass galaxies were moving on radial orbits, they could not retain their gas until reaching the pericenter, as their shallower gravitational potential would result in earlier stripping. Furthermore, radial orbits imply small pericentric distances, which is inconsistent with the large projected distances of these galaxies from the cluster center. These points effectively rule out radial orbits for at least a fraction of low-mass galaxies with observed tails.  

For low-mass galaxies on tangential orbits, the detectability of tails may be subject to observational biases. Gradual stripping over longer timescales would produce filaments of gas more extended along the orbit. Given the lower initial gas mass, such tails would be less dense and consist of more sparse clumps or ``fireballs'', making them more difficult to detect.

Ram pressure is responsible for compressing the ISM~\citep{2020MNRAS.497.4145T, 2020ApJ...901...95C}, which can trigger bursts of star formation~\citep{2022ApJ...941...77R}. However, the efficiency of ram-pressure-induced starbursts strongly depends on galaxy properties such as orbital parameters and disk orientation relative to the velocity vector. These factors can either enhance or suppress the starburst~\citep{2020MNRAS.494.1114S, 2021IJAA...11...95H}. Strong starbursts are typically observed in galaxies with predominantly edge-on orientations relative to their motion, whereas face-on galaxies generally do not experience significant enhancement in star formation.  

Another key factor is the galaxy orbit. Gas clouds in the ISM need sufficient time under compression to form stars before being fully stripped; in other words, the ram pressure must not increase too rapidly. From Eq.~\ref{eq:dPdt}, the growth of \(P_\mathrm{\rm ram}\) is proportional to the radial component of the galaxy velocity in the cluster. Therefore, ram-pressure-induced starbursts are inefficient for low-mass galaxies on radial orbits where $v_r \approx v$, which experience rapid and violent stripping. In contrast, an infalling galaxy with the same total orbital energy but higher angular momentum will have a smaller \(v_r\), making ram-pressure-induced star formation more efficient, since \(v_r\) is the only velocity component not contributing to the absolute value of angular momentum. Therefore, we expect that, in the low-mass regime, the morphological transformation from a gas-rich disk galaxy into a quiescent system via the post-starburst stage occurs most efficiently for galaxies on tangential orbits. 

Our cumulative distribution functions (CDFs) for both SFGs and PSGs indicate that galaxies of both these types tend to have higher angular momentum. However, as noted above, the structure of the CDFs differs between the two populations. Given our selection criteria for SFGs, which preferentially select low-mass galaxies, almost all of the selected SFGs are expected to be quenched in the future, even before reaching the cluster center. 
Currently, a substantial fraction (up to 40\%) of these SFGs despite being observed in a virial cone, are likely to be outside $R_\mathrm{vir}$ of the cluster~\citep*{2011MNRAS.416.2882M}. However, we expect that the vast majority of these objects are still close to the cluster, i.e. $r < 2-3 R_\mathrm{vir}$ and a gravitational potential of the cluster still dominates their motion, meaning that they will enter the cluster on a timescale of 1--2\,Gyr. 
Therefore, the SFG sample effectively probes the population of infalling, gas-rich dwarf galaxies that will be quenched within less than a billion years. In contrast, the PSG sample represents galaxies that have already undergone a triggered starburst followed by quenching. Ram-pressure is the most likely mechanism driving this transformation, as other processes, such as galaxy mergers, are less efficient in these systems due to higher velocity dispersions and lower galaxy densities in the cluster outskirts, where most PSGs reside. Thus, the combined study of these two samples allows us to trace the morphological transformation of galaxies within the same stellar mass bin via ram-pressure stripping. 

Conclusions about the tangentially of the orbits can be fully based on the interpretation of the PDF and CDF of $\hat{L}$ mostly in a case of a galaxy being far from the pericenter of the orbit given that, at pericenter, $\hat{L}$ would be always close to 0, since the angle between radius and velocity vectors is always around 90$^{\circ}$ regardless the total angular momentum of the orbit, i.e. even in a case of a nearly radial orbit. On the diagram of the permitted combinations of $r$ and $\hat{L}$, which we show on Fig.~\ref{fig:Lr_exmpl} this situation would correspond to $r \approx r_\mathrm{\perp}$, which does not constrain much $\hat{L}$ and allows for low $\hat{L}$ values. In such case a parameter, which defines the galaxy orbit is a absolute value of a pericentric distance, which is unambiguously bind to the absolute value of a total angular momentum and total orbital energy. For the identified PSGs, even projected distances to the cluster center are very large: on average they are close to $0.5 \,R_\mathrm{vir}$, indicating that a 3D distance to the cluster center and, therefore, a total angular momentum, should indicate tangential orbits.

Not all low-mass galaxies, which infall on-to a cluster can experience a ram-pressure-induced starburst and pass through a PSG phase lasting up to a billion years. The CDFs for SFGs and PSGs reveal a clear distinction: PSGs tend to have higher \(\hat{L}\), whereas SFGs exhibit more bimodal distributions. For SFGs, one subpopulation behaves similarly to PSGs, with \(p(\hat{L})\) increasing rapidly for high \(\hat{L}\) and peaking near \(\hat{L}=1\). A second subpopulation shows \(p(\hat{L})\) rising quickly at low \(\hat{L}\) and reaching a maximum around \(\hat{L} \approx 0.4-0.5\), suggesting that these galaxies move on more radial orbits. Although a similar low-\(\hat{L}\) population exists among PSGs, it constitutes a much smaller fraction.  

This results, under assumption of conservation of key orbital parameters, such as total energy and angular momentum, imply that only low-mass, infalling galaxies with high angular momentum -- i.e., those moving on tangential orbits -- undergo ram-pressure-induced starbursts and pass through a PSG phase for extended periods, up to a full orbital period within the cluster (approximately 1~Gyr, according to fig. B1 of \citealt{2017MNRAS.471.4170T}).

Galaxies with high \(\hat{L}\) during the jellyfish stage can have tails oriented in directions other than away from the cluster center. This behavior is also observed in more massive galaxies and can also result from tangential orbits, rather than stripping occurring solely near the pericenter of the orbit.  

Low-mass galaxies on radial orbits also undergo ram-pressure stripping; however, in their case, ram-pressure-induced starbursts are inefficient at forming significant stellar mass. These systems evolve into quiescent galaxies with red colors on timescales shorter than 200~Myr, rapidly bypassing the PSG stage. Since their overall stellar mass is lower than that of PSGs, due to the absence of a ram-pressure-induced starburst, these galaxies transform into low-mass dwarfs. Coupled with the expansion of stellar orbits following rapid gas removal, their end state can be ultra-diffuse galaxies (UDGs) with the lowest stellar masses, which they reach within 2--3~Gyr after quenching~\citep{2021MNRAS.506.4516L}.

We expect that further advances in deep spectroscopic surveys, such as DESI, 4MOST, WEAVE, PFS, MOONS, will enable deeper and more systematic studies of the effect of galaxy properties, including their orbits, on the morphological transformation in the low-mass regime, which is crucial for our understanding of the numerically dominant galaxy population in clusters.

\begin{acknowledgements}
IC's research is supported by the Telescope Data Center at the Smithsonian Astrophysical Observatory, by the Smithsonian Institution FY25 combined call grant, and by the HST-GO-17519 grant from the Space Telescope Science Institute.
AS's research was supported by a Lomonosov Moscow State University scholarship for young researchers and students.
\end{acknowledgements}

\bibliographystyle{aa}
\bibliography{PSG_Coma_DESI}

\begin{appendix}

\section{Robustness of the PDF of $\hat L$ on the model parameters}
\label{sec:Lstab}

Our estimates of the possible values of the normalized angular momentum were derived under specific assumptions about the Coma cluster galaxy and mass distribution parameters, as well as additional assumptions, such as zero total orbital energy.

In most cases, the uncertainties in these parameters do not significantly affect our modeling or conclusions. For example, the choice of the cluster center in a case of Coma can be ambiguous -- it may be identified with either of the two cD galaxies or with a point midway between them. However, since most of the selected PSGs and SFGs are located in the plane of the sky at distances more than ten times the separation between the two cDs, the impact of this assumption is negligible. 

Our modeling results are also dependent on the properties of Coma dark matter halo. Dynamical modeling done by \citet{2003MNRAS.343..401L} indicates an unusually high concentration of the dark matter halo of Coma of $c=9.4$, however, going down to typical values of $c=4$ does not affect the results since it leads to any noticeable changes only at large radii, exceeding $3-4R_\mathrm{vir}$, which do not have a significant contribution in the resulting PDF due to low galaxy density and, hence, low $f(r \mid r_\mathrm{\perp})$.

Below, we examine the robustness of our results with respect to variations in the remaining model parameters.

\subsection{Dependence on $v_\mathrm{sys}$ of the cluster}

The central region of the Coma cluster is dominated by two massive elliptical galaxies, NGC~4874 and NGC~4889, with line-of-sight velocities $v_\mathrm{\rm LOS}$ of 7183\footnote{\url{https://rcsed2.voxastro.org/data/galaxy/30609}}~${\rm
 km\,s^{-1}}$ and 6466\footnote{\url{https://rcsed2.voxastro.org/data/galaxy/30679}}~${\rm
 km\,s^{-1}}$, respectively. This substantial velocity difference suggests a recent merger between Coma and another massive cluster, introducing uncertainty in the systemic velocity of the cluster at the level of $\sim 200$–$300~{\rm km,s^{-1}}$.

To evaluate the impact of this uncertainty on the estimate of $\hat{L}$, we derived the dependence of $L$ on variations in $v_z$:

\begin{equation}
\frac{dL}{dv_z} \propto \sqrt{\frac{1}{L^2} - 1}.
\end{equation}

This relation shows that the effect is most significant at low $\hat{L}$, whereas the identified PSGs and SFGs generally exhibit higher probabilities of having large $\hat{L}$ values.

\subsection{Dependence on deviations from the $E = 0$ assumption.}

The presence of galaxy peculiar velocities ($v_\mathrm{\rm pec}$) can lead to deviations of the total orbital energy from the assumed value of zero. In fact, these $v_\mathrm{\rm pec}$ are responsible for producing non-zero angular momentum. However, their impact on $E$ and $\hat{L}$ differs significantly.

Since $\hat{L} \propto R\,v_\mathrm{pec}$ where $R$ is the distance from the cluster center, the effect of $v_\mathrm{\rm pec}$ on $\hat{L}$ is linear and therefore substantial. This effect becomes even stronger in a massive cluster, particularly when a galaxy is treated as a member of an infalling system at larger $R$, which naturally corresponds to larger $\hat{L}$.

In contrast, the kinetic energy scales with the square of the velocity, so the statistical contribution of $v_\mathrm{\rm pec}$ to the total energy is quadratic:

\begin{equation}
\frac{\delta E}{ U} \propto \left(\frac{v_\mathrm{\rm pec}}{v_\mathrm{\rm ff}}\right)^2.
\end{equation}

Given that in the field and filaments $v_\mathrm{\rm pec}$ typically does not exceed 200--300~${\rm km\,s^{-1}}$, while the free-fall velocity $v_\mathrm{\rm ff}$ at $R_\mathrm{\rm v}$ is of order 2000--3000~${\rm km\,s^{-1}}$, the resulting effect on $E$ is negligible.
\end{appendix}
\end{document}